\documentstyle [12pt] {article}
\def\references{\subsection*{REFERENCES}
\bgroup\parindent=0pt\parskip=\itemsep
\def\refpar{\par\hangindent=1.2em\hangafter=1}}
\def\endreferences{\refpar\egroup}

\def\@biblabel#1{\relax}
\def\@cite#1#2{#1\if@tempswa , #2\fi}
\def\reference{\relax\refpar}
\def\@citex[#1]#2{\if@filesw\immediate\write\@auxout{\string\citation{#2}}\fi
\def\@citea{}\@cite{\@for\@citeb:=#2\do
{\@citea\def\@citea{,\penalty\@m\ }\@ifundefined
{b@\@citeb}{\@warning
{Citation `\@citeb' on page \thepage \space undefined}}%
{\csname b@\@citeb\endcsname}}}{#1}}
\def\gsimeq
{\hbox{\raise0.5ex\hbox{$>\lower1.06ex\hbox{$\kern-1.07em{\sim}$}$}}}
\def\lsimeq
{\hbox{\raise0.5ex\hbox{$<\lower1.06ex\hbox{$\kern-1.07em{\sim}$}$}}}

\def\ergcms{ergs cm$^{-2}$ s$^{-1}$}
\def\ergs{ergs s$^{-1}$}
\def\pn{\par\noindent}

\begin{document}
\vglue0.5cm\noindent
\begin{center}
\vglue1.0cm\noindent
{\Large\bf The contribution of AGNs to the
\vglue0.3cm\noindent
 X-ray background}
\vglue1.0cm\noindent
{\large A. Comastri$^1$,
G. Setti$^{2,3,4}$, G. Zamorani$^{5,3}$, G. Hasinger$^{1,6}$}
\end{center}
\vglue2.0cm\noindent
{\sl $^1$Max Planck Institut f\"ur extraterrestrische Physik, Postfach 1603
 D--85740, Garching bei M\"unchen, FRG}
\vglue0.2cm\noindent
{\sl $^2$Dipartimento di Astronomia, Universit\'a di Bologna, via Zamboni 33,
I--40126, Bologna, Italy}
\vglue0.2cm\noindent
{\sl $^3$Istituto di Radioastronomia del CNR, via Gobetti 101,
I--40129 Bologna, Italy}
\vglue0.2cm\noindent
{\sl $^4$European Southern Observatory,
Karl--Schwarzschild--Strasse--2, D--85748, Garching bei
M\"unchen, FRG}
\vglue0.2cm\noindent
{\sl $^5$Osservatorio Astronomico di Bologna, via Zamboni 33, I--40126
Bologna, Italy}
\vglue0.2cm\noindent
{\sl $^6$Astrophysikalisches Institut, An der Sternwarte 16, D--14482
Potsdam, FRG}
\vglue1.5cm\noindent

 \centerline{\bf Astronomy and Astrophysics: in press}

\newpage\noindent

  \centerline {\Large\bf ABSTRACT}

\vglue2.0cm
We report the results of a detailed analysis of the contribution of various
classes of AGNs (Seyfert galaxies and quasars) to the extragalactic X-ray
background (XRB).
The model is based on the unification schemes of AGNs,
on their related X-ray spectral properties in the light of recent observational
results and on the X-ray luminosity function derived by Boyle et al. (1993).
The integrated emission from AGNs, when folded with an appropriate cosmological
evolution law, can provide a good fit to the XRB over a wide energy range,
from several to $\sim$ 100 keV, while it contributes only about 74\% of the
ROSAT soft XRB.
The baseline model predictions have been checked against all available
observational constraints
from both hard and soft X-ray surveys (counts, redshift distributions
and average X-ray source spectral properties).
\vglue1.0cm\noindent
\vglue1.0cm\noindent
{\bf Thesaurus:} 13.25.3 -- 12.04.2 -- 11.14.1 -- 11.17.3


\newpage\noindent

  \section {INTRODUCTION}

\par
The problem of the origin of the extragalactic X-ray background
(hereafter XRB) has attracted much renewed attention in recent years
following in particular the results obtained with the X-ray satellites
GINGA and ROSAT
(Fabian \& Barcons, 1992 and Zamorani 1994 for recent reviews).
\par
An analytical fit to the spectrum above $\sim 3$ keV, where most
of the energy intensity resides, has been recently given by
Gruber (1992) in the energy range $\sim 3-60$ keV:

$$ I(E) = 7.877 E^{-0.29} exp(-E/41.13) ~~~ keV\ cm^{-2}\ s^{-1}\ sr^{-1}\
 keV^{-1}      \eqno (1)  $$

\pn
and with two power laws from 60 keV to $\sim$ 6 MeV:

$$ I(E) = 1652 E^{-2.0} + 1.754 E^{-0.7} keV\ cm^{-2}\ s^{-1}\ sr^{-1}\
 keV^{-1}.      \eqno (2)  $$

\pn
It is useful to remember that
in the 3-20 keV energy range the XRB spectrum is well approximated
by a flat power law with spectral index $\alpha \simeq 0.4$.
Below 3 keV, in the soft X-ray band ($\sim 0.1-2.0$ keV), the XRB spectrum
measured by ROSAT is significantly steeper than the extrapolation
of the high energy best fit.
The bulk of the emission in the softest energy range,
$\sim 0.1-0.5$ keV, is of galactic origin probably due to a
local bubble of gas with a temperature of about $10^6\ K$,
while in the 0.5-2.0 keV band, where the galactic contribution should
be less important, the XRB spectrum can be well
approximated by a power law with an energy index $\alpha = 1.0 \pm 0.2$ and
a normalization
of $13.6 \pm 0.2$ keV cm$^{-2}$ s$^{-1}$ sr$^{-1}$ keV$^{-1}$ at 1 keV
(Hasinger 1992).
Similar results, but affected by larger errors, have been obtained
from the analysis of a shallower field (Barber \& Warwick 1994).
However in the $\sim 0.5-0.9$ keV passband (M band)
about 40\% of the flux may be due to the
emission of a hot gas with a temperature of
$\sim 2.2 \times 10^6\ K$,
although the origin of such component
is not yet understood (Hasinger 1992; Wang \& McCray 1993).
The remainder of the flux in the M-band and essentially all the flux above
$\sim 0.9$ keV are likely to be of extragalactic origin.

\par
The absence of any deviation from a pure blackbody spectrum of the cosmic
microwave background radiation as measured by the FIRAS instrument
on board COBE has set a very stringent upper limit ($\sim 10^{-4}$)
on the contribution of a diffuse hot intergalactic gas to the hard XRB
(Wright et al. 1994), leaving the alternative hypothesis on the origin of the
XRB as due to the sum of discrete sources the only attractive one.
Among the extragalactic sources, the Active Galactic Nuclei (Quasars and
Seyfert galaxies, hereafter AGN)
are known to be the strongest X-ray emitters and
have been considered to be the prime candidates able
to satisfy the energy requirements of the XRB.
However the fact that their X-ray spectra in the 2-10 keV energy interval are
characterized by an average slope of $\alpha \sim 0.7$
(Mushotzky 1984; Turner \& Pounds 1989) much steeper than that of the XRB,
has been, for a long time, the main problem for models
in which most of the XRB is due to AGNs.
\par
The GINGA discovery of a flattening of the hard X-ray spectra
of Sy 1 galaxies at $ > 8$ keV, explained either as a
partial coverage of the sources (Matsuoka et al. 1990) or as a
reprocessed flux from a relatively cold plasma (Pounds et. al 1990),
has led to several new attempts to fit the overall XRB spectrum
as a superposition of AGN spectra (Morisawa et al. 1990;
Fabian et al. 1990; Rogers \& Field 1991; Terasawa 1991).
A key feature of most of these models is the assumption of a strongly
enhanced reflected component, much
larger than observed in Sy 1 galaxies.
It has later been shown that these models do not meet
the constraints imposed by the source
counts and/or source spectra in the soft X-ray band (Setti 1992,
Comastri 1992).
In addition, Zdziarski et al. (1993a) have argued that these
models do not fit the position and the width of the peak of the
XRB spectrum. This conclusion has been refuted in the case
of the Fabian et al. model (Reynolds et al. 1994).
\par
A very good fit to the XRB spectrum in the energy range $\sim 2-100$ keV
has been obtained in a subsequent paper (Zdziarski et al. 1993b).
In order to be consistent with the GINGA observations and the recent OSSE
results ($\sim 50-300$ keV),
the mean AGN X-ray spectrum has been described by a thermal Comptonization
model with a temperature of $\sim 40$ keV
and an optical thickness of a few ($\tau_T \sim 3$), leading to a power
law with $\alpha = 0.7$ in the 2-10 keV energy interval with about half
of the flux reflected by a cold surface. The fit to the XRB has been achieved
by integrating up to a redshift cutoff $z_{max} = 4$ with the
cosmological evolution law which best fit the results of
Boyle et al. (1993). This model has not been checked for
consistency
with the soft X-ray source counts, but it is likely that the
predicted soft X-ray source spectra are too flat compared with
those actually observed in the ROSAT surveys.
\par
A different approach, based on the X-ray properties of unified
schemes of AGNs, has been suggested by Setti \& Woltjer
(1989).
In the simplest version of unified models (Antonucci 1993, for a
review) the different observed properties of various
types of AGNs (e.g. Sy 1 versus Sy 2) are entirely due
to the different orientation of the assumed absorbing molecular
torus with respect the line of sight.
For reasonable values of tori masses and sizes, the nuclear
emission can be absorbed up to $\sim 20-30$ keV depending on the
torus column density and geometry.
The existence of the torus has now been confirmed in the most
dramatic way by direct imaging with HST (Jaffe et al. 1993).
With simple hypotheses on the source spectra, assumed to have the
``canonical" $\alpha = 0.7$ slope, on the percentage distribution
of absorption cutoffs in the observed source population
and on the source cosmological evolution properties, Setti \& Woltjer
demonstrated that it is possible to reproduce the flat slope
of the XRB spectrum below $\sim 20$ keV, thus removing
the main objection to the interpretation of the hard XRB in terms
of the observed properties of AGNs.
The required number of absorbed sources (Sy 2) is approximately
equal to that of unabsorbed ones.
A more sophisticated model has been computed by
Madau et al. (1993). For two different assumptions on the source spectra
and evolution, and by requiring a fine tuning of
the optical thickness ($\tau_T = 2-3$) of the molecular torus,
they obtain excellent fits to the XRB spectrum up to $\sim 100$ keV.
A set of good fits to the XRB spectrum have been also obtained by Matt and
Fabian (1994) assuming a broken power law for the source spectra and
a distribution of absorbed sources with column densities ranging
from $10^{23}$ cm$^{-2}$ to $10^{25}$ cm$^{-2}$. Including the contribution
of the iron absorption edge and emission line on the source spectra, some
features in the XRB spectrum between 2 and 3 keV are expected.
\par
It is clear from the results discussed above that using the most recent
AGN spectral data it is possible to obtain good fits to the XRB spectrum.
However, on this basis only, it is difficult, if not impossible, to
discriminate between
the proposed models: almost equally good fits can be obtained with
significantly
different assumptions on the source spectra and evolution.
As pointed out by Zamorani (1994), a good fit to the XRB spectrum
{\bf is not} a sufficient condition to conclude that the problem of the
production of the XRB is solved.
\par
In this paper we shall discuss a self-consistent AGN model for the
synthesis of the XRB in the framework of the simplest AGN unified
scheme mentioned above, taking into account
the observational constraints provided by the
observed source number counts in the
soft (ROSAT and {\it Einstein}) and hard (HEAO-1 A-2 and GINGA) X-ray energy
bands, the respective redshift distributions and spectral
characteristics and the
relative distributions of different types of sources (i.e. Type 2 vs. Type 1)
as a function of the limiting flux and energy range.
\par
In $\S$2 we summarize the observational constraints on the X-ray spectral
and cosmological properties of AGNs provided by the various soft and hard
X-ray surveys.
In $\S$3 we discuss our basic model and compare its predictions with
all the available observational constraints.
The main results are discussed in $\S$4 and summarized in $\S$5.
\par
Throughout this paper
the adopted values for the Hubble constant and the cosmological deceleration
parameter are $H_0 = 50$\ km s$^{-1}$\ Mpc$^{-1}$ and $q_0 = 0$.


 \section {AGN X-RAY SPECTRA, LUMINOSITY FUNCTION AND COSMOLOGICAL EVOLUTION}

 \subsection{AGN broad band X-ray spectra}

Large samples of AGNs have been observed in the soft X-ray band with the
{\it Einstein} IPC in the energy range $\sim 0.2-4.0$ keV and with the ROSAT
PSPC in the energy range $\sim 0.1-2.4$ keV.
The {\it Einstein} IPC spectra of Sy 1 galaxies and quasars cover a wide
range of power-law spectral indices with a mean $\alpha \sim 1.0$
for the radio-quiet, and a mean $\alpha \sim 0.5$
for the radio-loud objects, with a large fraction of the objects showing an
upturn of the spectrum
below $\sim 0.5$ keV (Wilkes \& Elvis 1987; Kruper, Urry and Canizares 1990).
Since the contribution of the radio-loud objects to the soft XRB
at 2 keV is only a few percentage points (Della Ceca et al. 1994), for
simplicity we shall not treat them as a separate population.
\par
The analysis of two samples detected in the
ROSAT all sky survey shows a gradual flattening of
the mean spectral slope with increasing redshift from about
$\alpha \sim 1.5$ for nearby AGNs to $\alpha$ $\sim 0.8-0.9$
at redshifts $\sim 2.0-2.5$ (Walter \& Fink 1993, Schartel 1994).
The most likely interpretation of this effect
is a flattening toward the medium energies
of the intrinsic spectrum redshifted in the ROSAT band.
\par
In the $\sim 0.1-2.0$ keV energy range a power law fit to the summed
spectra of the sources detected in ROSAT deep surveys fields
suggests an energy index $\alpha \sim 1.0$
(Hasinger et al. 1993, hereafter H93).
\par
The observations from the GINGA satellite of a
large sample of
bright AGNs, mainly Sy 1 galaxies (Nandra 1991), have revealed several
new spectral features in the energy range $\sim 2-20$ keV.
In particular, an emission line complex near 6.5 keV along with an iron
K-absorption edge around 7-8 keV and a spectral ``hump" above $\sim 10$ keV
are common properties of the sample.
All these features are widely interpreted as the re-processing
(``reflection") of the primary X-ray radiation by relatively cold matter in
the proximity of the central source, probably associated with the accretion
disk (Pounds et al. 1990).
The percentage of the flux which is observed in the reflected component is
approximately equal to that present in the direct flux, implying that the
reflecting matter is consistent with an accretion disk subtending
about $2\pi$ sr at the X-ray source.
An important result of this ``reflection model" fit is that the
resulting slope of the intrinsic power
law spectrum ($\alpha \sim 0.9$) is significantly steeper than the
observed ``canonical" slope $\alpha \sim 0.7$
(Mushotzky 1984; Turner \& Pounds 1989).
As pointed out by Zdziarski et al. (1993b) the reflection component
below 10 keV is very weak and it seems difficult to explain the
flattening of the average 2-10 keV power law, by $\Delta \alpha \sim 0.2$,
as only due to the contribution of the reflected flux.
However, the results of a careful analysis of 61 high quality spectra of
Sy 1 galaxies observed by GINGA (Nandra \& Pounds 1994) indicate that
the observed ``canonical" slope of $\alpha \sim 0.7$ is probably due to the
combined effects of reflection and warm absorber features on the underlying
mean X-ray continuum with slope $\alpha \sim 0.9-1.0$.
The signature of highly ionized material (O VII - O VIII) at
$\sim 0.7-0.8$ keV, the
so-called warm absorber, has been revealed in few high signal-to-noise
ROSAT observations of AGNs (Nandra \& Pounds 1992;
Nandra et al. 1993; Turner et al. 1993a; Fiore et al. 1993).
\par
Recent analyses of EXOSAT and GINGA observations of high luminosity AGNs
(Comastri et al. 1992; Williams et al. 1992) have shown that the
distribution of the power law energy index of QSOs has a mean value
$\alpha \sim 0.9$, steeper than the one observed for lower luminosity
AGNs.
While the Sy 1 galaxies are well described by a power law plus a
disk reflection model, the higher luminosity sources do not show
evidence of the reflected component.
\par
The unified scheme scenario is supported by GINGA observations of a
relatively large
sample of Sy 2 galaxies (Awaki et al. 1991), where most of the
detected sources (14 out of 28 observed) show evidence of intrinsic absorption
with column densities in the range $10^{22} - 10^{24}$ cm$^{-2}$.
The non--detection in hard X-rays of other optically selected Sy 2
may be an indication of even larger column densities, of the order of
$\sim 10^{24} - 10^{25}$ cm$^{-2}$ with solar composition.
It should also be noted that intrinsic absorption, with column densities
in the range $N_H \sim 10^{21-23}$ cm$^{-2}$,
has been observed in the X-ray spectra of Seyfert galaxies of intermediate
types (Turner \& Pounds 1989; Nandra \& Pounds 1994), suggesting the
existence of a continuous distribution in the absorption column densities
in the Seyfert population.
\par
Soft X-rays were also detected from some Sy 2 galaxies (Kriss,
Ca\-ni\-za\-res \& Ricker 1980), but the implied X-ray luminosities are
in general much smaller (two or more orders of magnitude) than those of
Sy 1 galaxies of comparable optical magnitude.
ROSAT X-ray spectra of Sy 2 galaxies have been recently published
(Turner et al. 1993b; Mulchaey et al. 1993). They generally
show substantial photoelectric absorption with a range of column densities
up to several times $10^{22}$ cm$^{-2}$ in agreement with GINGA observations.
In some cases the observed X-ray fluxes are greater than expected
by simple extrapolations from the high energy data plus a uniform absorber,
suggesting a different
origin for the soft component. The soft X-ray  emission has been
interpreted, in the framework of unified schemes, as electron scattered
nuclear radiation.
At least in the case of NGC 1068 the soft flux is known to arise from
an extended starburst region (Wilson et al. 1992).
\par
There is a growing observational evidence that, as suggested
by Barthel (1989), radio-loud quasars and strong radio galaxies
can also be unified in a way similar to the Seyferts.
X-ray observations of strong radio galaxies tend to support
this unified picture (Koyama 1992; Allen \& Fabian 1992).
In addition, the discovery of absorption ($N_H \sim 10^{22}$ cm$^{-2}$)
in the ROSAT spectra of high-redshift quasars (Elvis et al. 1994)
reverses the trend
for the most luminous AGN to show the least X-ray absorption.
It should also be borne in mind that the possible existence
of an obscured population of intrinsically luminous radio-quiet quasars
is still uncertain, although Sanders et al. (1989) have suggested that
the IRAS ultra-luminous
galaxies might be the obscured counterpart of luminous quasars.
\par
Until recently, the spectral properties of AGNs at higher energies
were essentially limited to the early balloon observations of
bright nearby AGNs (i.e. 3C273; Bezler et al. 1984).
The OSSE instrument on the
{\it Compton Gamma-Ray Observatory} has now detected hard X-ray emission
($E > 50$ keV) from numerous Seyfert galaxies (Johnson et al. 1994).
The OSSE spectrum of NGC 4151 (Maisack et al. 1993)
is very soft and can be described by a broken power law
with $\alpha = 1.1\pm0.3$ below 100 keV and $\alpha = 2.4\pm0.4$
above the break energy, or by a thermal Comptonization model
with a temperature of $\sim 40$ keV.
This is consistent with earlier observations obtained
by the SIGMA telescope onboard GRANAT (Jourdain et al. 1992).
Somewhat different results have been obtained from the OSSE observations
of the Sy 1 galaxy IC 4329A, which is likely to be more
representative of the Seyfert population than NGC 4151 (Madejski et al. 1994).
Joint fits of the ROSAT, GINGA and OSSE data
are well modeled using an exponentially cut-off power law continuum
with an e-folding energy $240\ \lsimeq\ E\ \lsimeq \ 900$ keV
plus reflection (Madejski et al. 1994).
\par
The existence of a high energy cutoff is also clearly indicated by
the analysis (Johnson et al. 1994) of the average OSSE spectrum
from a sample of 15 Seyfert galaxies.
Moreover,
the non--detection of a hard X-ray selected sample of Seyfert galaxies
at energies $ > 100$ MeV by EGRET (Lin et al. 1993) is consistent with
the break in the spectrum at a few hundreds keV being
a common spectral feature of the Seyfert population.
\par
The few radio-loud quasars which have been detected by OSSE
(Johnson et al. 1994)
generally show harder spectral indices than Seyfert galaxies,
consistently with
the EGRET detection at energies greater than 100 MeV of more than 20
core dominated radio quasars (Kurfess 1994).
As pointed out by Padovani et al. (1993) and Setti \& Woltjer (1994)
the radio-loud objects detected by EGRET
may well provide a relevant fraction of the $> 100$ MeV
$\gamma$-ray background.

 \subsection {AGN X-ray luminosity function and evolution}

Our knowledge of the AGN X-ray luminosity function and evolution
in the soft X-ray band has been greatly expanded with
the almost complete optical identifications of the Extended Medium Sensitivity
Survey (EMSS) made in the 0.3-3.5 keV passband (Gioia et al. 1990, Stocke
et al. 1991) and the ROSAT quasar survey in the 0.5-2.0 keV energy range
(Shanks et al. 1991).
Models for the AGN X-ray luminosity function (XLF) have been constructed
from the EMSS AGNs (about 450 objects) by Maccacaro et al. (1991,
M91) and from
a combination of the EMSS AGNs with 42 new QSOs from two deep ROSAT
pointings by Boyle et al. (1993; B93).
In the framework of a pure luminosity evolution model
the best XLF (cfr. B93) is represented by two power laws of the form:

 $$ \rho(L,z=0) = K_1 L_{44}^{-\gamma_1} ~~~~~~~~~for~~~ L_x < L_B \eqno (3) $$

 $$ \rho(L,z=0) = K_2 L_{44}^{-\gamma_2} ~~~~~~~~~for~~~ L_x > L_B \eqno (4) $$

\noindent
where $L_{44}$ is the 0.3-3.5 keV X-ray luminosity expressed in units of
$10^{44}$ \ergs. The evolution of the XLF is parameterised as a power
law in $(1 + z)$:

 $$ L_X (z) = L_X (0) \times (1 + z)^{\beta}                  \eqno(5) $$

\pn
Assuming $q_0 = 0$ and an X-ray spectral index $\alpha = 1$ an acceptable
fit to the combined EMSS/ROSAT data has been obtained with the following
parameter values for the XLF:
$\gamma_1 = 1.7 \pm 0.2$, $\gamma_2 = 3.4 \pm 0.1$,
$K_1 = K_2 L_{B}^{\gamma_1 - \gamma_2} = 5.7 \times 10^{-7}\ Mpc^{-3}\
(10^{44}\ ergs\ s^{-1})^{\gamma_1 - 1} $,
log $L_B = 43.84 \pm 0.1$ and $\beta = 2.75 \pm 0.1$.
These parameters have been obtained over the full luminosity-redshift
range $10^{42} < L_X < 10^{48}$ \ergs $\ $  and $0 < z < 3$.
The inclusion of a redshift cutoff above which the evolution
stops ($z_{cut} \sim 2$), although not required by the data for
$q_0 = 0$, improves the acceptability of the model.
\par
Because of the soft X-ray selection the parameters of the XLF and evolution
derived from the {\it Einstein} and ROSAT data can be considered to
adequately represent the population of unabsorbed AGNs.

  \section {A SELF-CONSISTENT APPROACH TO THE SYNTHESIS OF THE XRB}

This section is devoted to a detailed discussion of an AGN model
for the XRB in the framework of the X-ray unified scheme.
\par
The underlying assumption is that the shape of the XLF and evolution
of absorbed objects is the same as that of the unabsorbed ones
which entails that the $N_H$ distribution is independent from the source
luminosity.
The luminosity range chosen for the integration is
$10^{42} < L_X(0) < 10^{47}$ \ergs $\ $.
We have further assumed $L_B$ (i.e. the break luminosity in the XLF
at redshift zero) as the dividing luminosity
between the low luminosity unabsorbed objects (i.e. Sy 1 with the spectrum
of Eqs. 6,8) and the high luminosity unabsorbed objects
(i.e. quasars with the spectrum of Eqs. 6,7).
We have also assumed that the gaseous matter responsible for the
absorption has solar chemical composition independent of the
object redshift.
Finally, it should be noted that we did not try to reproduce
the full intensity and spectrum of the soft (below a few keV)
XRB. In fact it is already known that in the soft X-ray band other
classes of sources (e.g. galaxies and clusters of galaxies) provide
a non--negligible contribution to the XRB.
Also, due to the large number of free parameters involved in the
calculations we did not try a global best fit of the whole set of
observational constraints, but we have followed the approach described below.

  \subsection{The baseline model and the fit to the XRB}

In this model the parameters for
the broad band X-ray spectrum and cosmological evolution of different types
of AGNs have been chosen in order to obtain a good fit to the overall
set of observational constraints.
All the assumed parameters are consistent, within the errors, with those
suggested by the present available observations.
The adopted spectral shape of the different types of AGNs
(cfr. $\S$2) can be written as follows:

$$  F(Quasars,Sy\ 1) \propto E^{-\alpha_{s}} ~~~~~~~~ for ~~ E < 1.5 ~ keV
    \eqno(6) $$

$$  F(Quasars) \propto E^{-\alpha_h} \ exp \left(-{{E} \over {E_c}} \right)
    ~~~~~~~ for ~~ E > 1.5 ~ keV  \eqno (7) $$

$$  F(Sy\ 1) \propto E^{-\alpha_h} \ exp \left(-{{E} \over {E_c}} \right)
[1 + f_r A(E)] ~~~~~ for ~~ E > 1.5 ~ keV  \eqno (8) $$

$$  F(Type\ 2) = F(Type\ 1) \times  e^{-\sigma_E N_H}           \eqno (9) $$

\pn
where $\alpha_{s} = 1.3$ is the energy index in the soft band,
$\alpha_h = 0.9$ is the index above 1.5 keV and $E_c = 320$ keV is the
e-folding energy of the exponential spectrum.
The here adopted $E_c$ value has been taken from the results
of the combined ROSAT-Ginga-OSSE fit to IC 4329A (model D in
Madejski et al. 1994).
It should be noted that this falls within the range of $E_c$ values
obtained by fitting our power law plus reflection spectrum to the
OSSE data points of the average Seyfert spectrum in Johnson et al. (1994).
For the angular dependence of the X-ray radiation
reprocessed by the accretion disc, $f_r (\theta)$, we have adopted the
analytical approximation given by Ghisellini et al. (1994).
$A(E)$ is the reflectivity of the cold gas
(White, Lightman \& Zdziarski 1988; Lightman \& White 1988) and
$\sigma_E$ the photoelectric absorption cross section
assuming solar composition (Morrison \&
McCammon 1983).
As required by the unified scheme, the intrinsic X-ray spectrum of
type 2 AGNs has been assumed to be the same as the spectrum of type 1 objects,
modified by absorption effects, with
$N_H$ the column density of the molecular torus in units of atoms cm$^{-2}$.
Assuming $\theta_0 = 45^o$ for the torus half-opening angle
(Barthel 1989; Awaki et al. 1991; Goodrich et al. 1994)
the average value of $f_r$, weighted over the solid angle,
for the unabsorbed ($0 < \theta < \theta_0$) type 1 objects
is 1.29, while it is 0.88 for type 2 objects
($\theta_0 < \theta < 90^o$).

\par
The adopted XLF and evolution are those described in $\S$ 2.2
with one minor modification. The best fit evolutionary parameter
$\beta = 2.75\pm 0.1$ has been obtained by B93 under the
assumption of a single population of unabsorbed objects.
The possible presence in the sample of a fraction of
absorbed objects, which in the soft band are more easily detected
at high redshift, makes the derived luminosity evolution higher than the
real one. For this reason we have adopted $\beta = 2.6$ in our baseline
model. Moreover, since a lower evolution implies a higher normalization
of the local XLF, the $K_1$ value given by B93 is increased by 20\%.
As a compromise between the B93 suggestion ($z_{cut} \sim 2$) and the
most recent results on optically selected high redshift quasars
which suggest a higher redshift cutoff (Warren et al. 1994),
we have introduced a
redshift cutoff $z_{cut} = 2.25$. For $z > z_{cut}$ the XLF has
been assumed to be constant up to $z_{max} = 4.0$.
An additional important parameter of the model is the distribution
of absorbed objects as a function of $N_H$.
Such a distribution, normalized to the
number of unabsorbed AGNs ($N_H < 10^{21}$), has been estimated
by requiring acceptable fits to all the observational constraints
described in the following sub-section.
\par
The solid curve in Fig. 1a shows the fit to the XRB
resulting from our model, while the dotted and dot-dashed curves show the
contribution
of the various classes of AGNs as a function of intrinsic $N_H$.
For simplicity the absorbed objects ($N_H > 10^{21}$ cm$^{-2}$) have been
divided into four subclasses, one decade in $N_H$ wide,
up to $N_H = 10^{25}$ cm$^{-2}$. The number density of
objects in the four $N_H$ classes, normalized to the number of
unabsorbed AGNs is 0.35,1.10,2.30,1.65. The data for $E > 3$ keV are a
compilation of best results from various instruments, while the
soft data are from ROSAT (Hasinger 1992; solid lines).
Figure 1b shows the fractional difference between our model results
and the analytical approximation of the XRB in the energy range
3-100 keV (Eqs. 1 and 2). To help in judging the goodness of the fit,
we show in this figure a few typical measurement errors at various energies.
It is seen that, while for $E\ \gsimeq\ 5$ keV the computed intensity is
within two sigma from all the available data points,
at $E\ \lsimeq\ 5$ keV it falls significantly below the observed
data. This has been done purposely in order to accomodate minor
additional contribution from
other extragalactic sources such as clusters of galaxies,
normal and starburst galaxies (Setti 1990 and references therein).

  \subsection {Other Soft and Hard X-ray constraints}

The number of sources per steradian with flux $> S_{lim}$ in the energy
range $E_1 - E_2$
can be computed, for a given luminosity function $\rho(L,z)$, as:

 $$ N( > S_{lim}) = \left({{c} \over {H_0}} \right) \int_0^{z_{max}}
  \int_{max[L_{min} (S_{lim}), L_{min}]}^{(1+z)^{\beta} L_{max}(0)}
  {{D_L^2 (z)} \over {(1 + z)^3}} \rho (L,z) dL dz  \eqno (10) $$

\noindent
where:
$L_{min}(S_{lim}) = 4 \pi D_L^2 S_{lim} (E_1, E_2) / K(z, E_1, E_2) $,
$L_{min} = (1+z)^{\beta} \times 10^{42}$  \ergs, $D_L = {{c} \over {H_0}}
 z (1 + z/2)$
is the luminosity distance and $K(z, E_1, E_2)$ is the K-correction term.
\par
The expected counts for each class of sources of the ``baseline model"
(in the 0.5-2.0 keV passband) are presented in Fig. 2.
One can see that the total counts
are in good agreement with the observed EMSS AGN Log N - Log S (Della Ceca
et al. 1992)
converted from the 0.3-3.5 keV IPC to the ROSAT 0.5-2.0 keV
band assuming a spectral slope $\alpha = 1$.
The predicted surface density at the limiting flux $S_{lim} \sim 10^{-14}$
\ergcms $\ $ is in good agreement with the available optical
identifications of a number of deep ROSAT observations,
where the AGN content has been estimated to be in the range 60-85\%,
while at the faintest
flux limit $\sim 2.5 \times 10^{-15}$ \ergcms $\ $ it is close to
the observed one ($\sim 400$ objects per square degree).
The predicted AGN counts are also fully consistent with
the fluctuation analysis at even fainter fluxes.
\par
The redshift distribution in the energy range 0.3-3.5 keV
has been computed taking into account the sky coverage of the EMSS
(M91) as a function of the limiting flux, which in turns depends on the
source spectral shape. The flux limits quoted in M91 apply to the
unabsorbed AGNs, while they should be increased by about a factor 2
for sources with log $N_H = 21.5$ (Zamorani et al. 1988), the only relevant
contributors to the EMSS counts among the absorbed AGNs (Fig. 2).
As a consequence the derived distribution is largely dominated
by the unabsorbed AGNs ($\sim$ 97\%).
The results (Fig. 3) are in good agreement with the
EMSS AGN redshift distribution. The overprediction at low redshift
($z < 0.5$), which is statistically not significant, is likely to be due
to the adopted slope for the faint end of the XLF ($\gamma_1 = 1.7$) steeper,
but consistent within the errors, with the EMSS slope ($\gamma_1 \sim 1.4$,
Della Ceca et al. 1992).
\par
The predicted redshift distribution in the 0.5-2.0 keV energy range
has been computed, for the unabsorbed sources, down to the flux of
$\sim 10^{-14}$ \ergcms $\ $,
corresponding to the limiting flux of the already optically identified
sources in a number of ROSAT deep fields.
In analogy with the procedure described above, the flux limits for
the absorbed population has been
computed adopting the appropriate conversion factors for the
ROSAT 0.5-2.0 keV band.
The results (Fig. 4) are consistent with
the redshift distribution of the AGNs so far identified
in five deep ROSAT fields: Lockman hole, Marano field, NEP,
QSF1 and QSF3 (Hasinger 1994). Given the present size of the sample and
its incomplete identification,
the excess of predicted sources above $z \sim 2$ is statistically not
significant. If such an excess remains when a larger complete sample becomes
available, it may be necessary to lower the adopted value for $z_{cut}$.
\par
The average soft (0.5-2.0 keV) X-ray spectrum has been computed
for different flux limits ranging from $10^{-13}$ to $3 \times 10^{-15}$
\ergcms $\ $ (see Table 1). The mean slope shows a gradual flattening from
$\alpha \sim 1.0$ at the bright flux limit,
consistent with the average AGN spectrum of the {\it Einstein} EMSS sources
(Maccacaro et al. 1988), to $\alpha \sim 0.84$ at the faint end.
As shown in the last entry of the Table the average spectrum
becomes significantly harder at fluxes even fainter that the current
ROSAT limit. This trend is
consistent with what has been observed by H93 (see their Fig. 3).
The average spectrum of the ROSAT deep survey sources
in the Lockman hole ($<\alpha> = 0.96 \pm 0.11$; H93)
has been computed by summing the counts
of all the sources fainter than $4 \times 10^{-14}$ \ergcms $\ $
in the $\sim 0.1-2.0$ keV energy range.
The mean spectrum of the sources in our model, computed with the same
prescription of H93, is $\alpha = 0.99$, being thus in excellent agreement
with the observed one.
The gradual steepening of the mean soft X-ray spectrum toward
lower energies is also shown, for various limiting fluxes, in Table 1.
\par
In the hard (2-10 keV) X-ray band
the HEAO-1 A-2 all-sky survey (Piccinotti et al. 1982) provides
an important constraint.
Thirty AGNs, mostly Sy 1 galaxies, constitute a complete sample at the
flux limit of $2.74 \times 10^{-11}$ \ergcms $\ $ over 8.2 steradians
of the sky.
The predicted number of sources is lower than but consistent
(within $1 \sigma$) with the
Piccinotti et al. (1982) AGN surface density (Fig. 5).
Moreover it should be noted that the AGN counts
in the HEAO-1 A-2 survey are likely to be somewhat overestimated because of the
excess of sources in the local supercluster.
In fact, nine of the twelve sources with $z < 0.01$ are located
toward the central region
of the local supercluster, while only three are in the anticenter
direction. Moreover
all but one have supergalactic latitude less than $|30^0|$.
The excess of sources in the Piccinotti et al. sample
is thus estimated to be of the order of 20\%.
\par
Adding the contribution of clusters of galaxies,
the predicted counts are also consistent
with the region allowed by the GINGA (Hayashida 1990) and HEAO-1 A-2
(Shafer 1983) fluctuation analysis in the flux range
$\sim 10^{-13} - 10^{-11}$ \ergcms $\ $.
It should be noted that the fluctuation analysis includes
the contribution of all emitting X-ray sources (galactic and extragalactic).
The galactic contribution at $\sim 10^{-12}$ \ergcms $\ $
is estimated to be of the order of 10\% (Hayashida 1990).
\par
Also, the derived redshift (Fig. 6) and absorption (Fig. 7) distributions
of the AGNs at the flux limit of the Piccinotti sample
show a good agreement with those observed.
The X-ray spectra and absorbing column densities
for all the objects in this sample have been measured by EXOSAT (Turner \&
Pounds 1989) in the $\sim 0.1-10$ keV band and
by ROSAT (Schartel 1994) in the $\sim 0.1-2.4$ keV energy range.
The mean ratio between the hard (2-10 keV) and soft (0.5-4.5 keV)
X-ray fluxes $<S_H/S_S> = 2.05$ is consistent with the value
($1.95^{+0.40}_{-0.33}$) obtained by Franceschini et al. (1993).
The local X-ray volume emissivity of the baseline model in the
2-10 keV band is $3.8 \times 10^{38}$ \ergs\  Mpc$^{-3}$, fully
consistent with that derived by Miyaji et al. (1994) from
a cross-correlation analysis of the HEAO-1 A-2 all sky XRB maps
and the galaxies from the IRAS 2 Jy redshift survey sample
and the 0.7 Jy projected sample.

  \section {DISCUSSION}

\par
The baseline model described in the previous section
not only provides an excellent fit to the XRB from a several
keV up to $\sim$ 100 keV, but it is also consistent with
essentially {\bf all} the available observational constraints
in the soft and hard X--ray bands. In order to test
the sensitivity of the results to the various parameters
defining the model, we have run a number of different
models by modifying a few of the input parameters. In essentially
all the cases we have been able to reproduce acceptable
fits to the XRB, but worsening one or more of the additional
consistency checks described in Section 3.2. For example,
we have varied by 0.2 the slope of the faint end of the
XLF (within the 1$\sigma$ error quoted by B93). Using
a steeper slope ($\gamma_1$ = 1.9), the increased number of low luminosity
objects allows a better fit of the 2-10 keV AGN surface density (Piccinotti
et al. 1982), but the predicted number of low redshift ($z < 0.3$)
sources in the soft energy range increases significantly and
becomes inconsistent with the observed EMSS data.
Qualitatively the same effects are obtained if one extrapolates
the assumed XLF down to $L_X = 10^{41}$ \ergs .
Viceversa,
using a flatter slope ($\gamma_1$ = 1.5) allows an even better fit to the
low redshift distribution of EMSS AGNs, but underestimates significantly
the surface density of bright 2-10 keV AGNs.
We have also considered a different description of the X-ray
spectral properties, moving below 1.5 keV the low energy steepening
of the average spectrum. In particular, assuming a mean spectrum described by
a single power law ($\alpha = 0.9$) above
0.5 keV, we find that the mean ROSAT source spectrum
is significantly flatter, by $\Delta \alpha \sim 0.3-0.4$, than the
observed one.

The fit to the XRB does not depend critically on the precise form
of the mean source spectrum above 3 keV. For instance, good fits
have been obtained by assuming somewhat harder power laws
($\alpha \sim 0.7-0.8$) and no ``reflection" component.
As originally argued by Setti \& Woltjer (1989),
a dominant factor in this fit is represented by the distribution of absorbed
sources. This is why we have not attempted a more detailed
modeling of the source spectral characteristics: if on the one
hand it is true that not all Sy 1 galaxies present a ``reflection"
component, on the other hand we have not considered the additional
contribution of the ``warm absorber" component (cfr. $\S$ 2.1)
to our mean source spectrum.
Similarly, we have not considered possible modifications to
the observed spectra induced by Compton-thick tori. These
effects are important for the objects with the highest density
($N_H > 10^{24}$ cm$^{-2}$; Ghisellini, Haardt \& Matt 1994) and
their inclusion would not modify substantially any of our conclusions.

Likewise, above several tens keV the fit to the XRB spectrum is not
strongly dependent on the precise shape of the high energy spectrum
as long as an exponential cut-off or a break to a steeper slope is present.
In fact good fits have been also
obtained by adopting a broken power law model for the mean AGN spectrum
with a steep slope ($\alpha \sim 1.6$) above a break energy of 75 keV,
which provides a reasonable representation of the OSSE data points
for the average Seyfert spectrum (Johnson et al. 1994). The main
conclusions reached in this paper remain unchanged.
Due to the paucity of the spectral data above $\sim 50-100$ keV
both the average slope and the break energy are still very uncertain.

A detailed model for the XRB above $\sim$ 100 keV is beyond the purposes
of this paper and will be discussed elsewhere.

Obviously, given the large number of parameters which define
the model, we have not attempted a full exploration of the
entire allowable parameter space and therefore we can not
exclude that more complex modifications of the baseline
model could provide even better fits than those described
in the previous Section. However it is useful to discuss some
of the implications and testable predictions deriving from the
baseline model.

The fit to the XRB is within 6\% of the observed data from
6 keV up to 100 keV. In the range 3 to 6 keV the model
prediction is below the data by (5--10)\%. In this energy
range the contribution of clusters of galaxies to the XRB has been
estimated to be $\sim 5\%$ (Piccinotti et al. 1982) and can
therefore be easily accomodated in our model. On the contrary,
a dominant contribution from star-forming galaxies, as proposed by
Griffiths \& Padovani (1990), would be inconsistent with our model.

Below 2 keV a careful fit to the XRB spectrum has not been
attempted because other classes of extragalactic sources may contribute
significantly. In this band the XRB
is not totally accounted for by the AGN model discussed here.
At the flux limit of $3 \times 10^{-15}$ \ergcms $\ $ about 59\%
of the 1-2 keV XRB is already resolved into sources (this fraction rises to
at least 75\% on the basis of a P(D) analysis (H93)), to be compared with
about 50\% resulting from integration, at the same flux limit,
of our log N - log S relation for AGNs.
An extrapolation of our model to zero fluxes
accounts for $\sim 74\%$ of the 1-2 keV background.

Preliminary results of ASCA observations of the 1-10 keV XRB
spectrum (Gendreau et al. 1994) indicate that the high energy
power law ($\alpha \sim 0.4$) can be extrapolated down to $\sim$ 1 keV
with no evidence for a steepening around 2-3 keV, at variance with the
ROSAT observations.
If these results will be confirmed by future observations, the predicted
AGN contribution to the soft XRB of our model could be higher.

The AGN number counts predicted by B93
at the faintest ROSAT limit by integrating their XLF is lower by a factor
$\sim$ 1.8 than the total observed number of sources. The reason
for this difference is that at faint fluxes
the relative contribution of absorbed sources ($10^{21} \leq N_H
\leq 10^{23}$ cm$^{-2}$), not considered by B93, becomes increasingly more
important in our model (Fig. 2). We expect that the faint sources
with relatively hard X-ray spectra observed in
the Lockman Hole deep survey (see Fig. 3 of H93) will be
identified with absorbed AGNs at intermediate to high redshifts.

For four ROSAT deep fields a relevant fraction ($\sim 76\%$)
of sources have been already spectroscopically identified above a flux
of $\sim 10^{-14}$ \ergcms $\ $.
The results indicate that the stellar
fraction in the deep ROSAT surveys is at most 10\% and that,
among the extragalactic sources, unabsorbed type 1 AGN are
the dominant population ($\sim 60\%$), followed by
galaxies, clusters of galaxies, and BL Lac objects (Shanks et al. 1991;
B93; Zamorani 1994; Hasinger 1994). The number of clusters identified
in these surveys is still very low, also because their optical identification,
requiring spectroscopy on very faint galaxies, is difficult.
As a consequence, their contribution to the soft XRB
is uncertain, depending also on their still ill defined
cosmological evolution.
The integrated emission of
a diffuse relatively cool intergalactic medium in clusters of galaxies,
not precluded by the COBE results, could in principle complement the
AGN contribution so as to saturate the soft XRB (Burg, Cavaliere
\& Menci 1993).
\pn
The percentage of galaxies in the ROSAT deep surveys
at the limiting flux of $\sim 10^{-14}$ \ergcms $\ $ is relatively
small (5--10\%). On the basis of preliminary and still incomplete
optical identifications it has been suggested (Griffiths et al. 1994)
that early type and Narrow-Emission-Line galaxies
may become at least as numerous as the AGNs at even fainter fluxes
($5 \times 10^{-15} < S_x < 10^{-14}$ \ergcms $\ $).
This result, if confirmed by more complete optical identifications,
can still be accomodated in our model. In fact, it is seen from Fig. 2
that some room is left for fluxes around $10^{-15}$ \ergcms.
This is even more true if, as pointed out by Griffiths et al. (1994),
the presence of Sy 2 nuclei among their Narrow-Line galaxies
cannot be ruled out.

{}From the previous discussion on our model it is possible to obtain some
estimate on the number ratio between Sy 2 and Sy 1
galaxies. It should be stressed that the computed ratios between the
absorbed and unabsorbed populations can not be directly associated with the
number ratio between Sy 2 and Sy 1. In fact it is known that
there is a non negligible number of galaxies classified as Sy 1
which show substantial intrinsic
absorption, with column densities up to several times $10^{22}$ cm$^{-2}$
(Turner \& Pounds 1989, Nandra \& Pounds 1994). On the other hand, a number of
galaxies classified as Sy 2 show a column density smaller than
$10^{22}$ cm$^{-2}$. For this reason it is not easy to identify
the $N_H$ value which best separates the two classes of Seyfert galaxies.
However, if we tentatively consider as Sy 1 the unabsorbed objects and those
with relatively low intrinsic column densities
($N_H < 10^{22.0} - 10^{22.5}$ cm$^{-2}$), the resulting ratio between
Sy 2 and Sy 1 is in the range 2.4--3.7 in reasonable agreement
with the value of $2.3 \pm 0.7$
found by Huchra \& Burg (1992) for a complete sample of optically selected
Seyfert galaxies. From a finer subdivision, Osterbrock \& Martel (1993)
find that the ratio between Sy 1.8-1.9-2 and Sy 1-1.5 is 3.5-4.0.

One of the main results of our analysis is that,
contrary to the suggestion discussed by Franceschini et al. (1993),
there is no need for two distinct populations (soft and hard X-ray
selected) of AGNs with different spectral and evolutionary properties.
This can in principle be tested when deeper samples of hard X--ray
selected AGNs may become available: hard X-ray selected
sources should have the same evolutionary properties as the soft X-ray
selected ones.

One important assumption of the model, which needs to be tested
by future observations, is the existence of highly absorbed high
luminosity objects. Almost nothing
is known about the distribution of $N_H$ column density for high
luminosity radio quiet quasars. If such distribution turns out to be
different from that of lower luminosity objects, some
modifications to our baseline model would be required.
In addition, one hopes that future observations will permit
to extend our knowledge of the $z=0$ XLF below $10^{42}$ \ergs.


   \section {CONCLUSIONS}

\par
The main results of our paper can be summarized as follows:
\pn
{\bf a)}
In the framework of the AGN unified schemes
it is possible to construct a baseline model which
reproduces with good accuracy  the XRB spectrum over the
broad energy range $\sim 5 - 100$ keV and meets essentially all presently
know X-ray constraints on the source population and spectral
characteristics. The predicted AGN contribution to the ROSAT XRB
in the 1-2 keV band is about 74\%.
\pn
{\bf b)}
The key feature of the model is the existence of an AGN population
with absorbed X-ray spectra characterized by a distribution of intrinsic
column densities in the range $N_H = 10^{21}-10^{25}$ cm$^{-2}$.
{}From such a distribution we derive an estimate of the ratio
between Sy 2 and Sy 1 galaxies which is consistent with the observed
ratio in nearby optically selected complete samples.
\pn
{\bf c)}
The spatial distribution and cosmological evolution of both absorbed and
unabsorbed AGNs can be essentially described
by the XLF parameters derived by B93, provided
that the strong power law evolution shows a significant decline
or cut-off at a redshift $z = 2.0-2.5$.
\pn
{\bf d)}
The model calculations do not strongly depend on the details of the
spectral slope, and acceptable fits to all the observational data
can be obtained as long as an exponential cut-off or a break to a
steeper slope is present in the average AGN spectra above $\sim 70$ keV.

   \section {ACKNOWLEDGEMENTS}
We thank the referee, A.C. Fabian, for his constructive comments
and for pointing out that results similar to ours have been obtained
in a recent work by Madau, Ghisellini \& Fabian (1994, MNRAS, in press).
We appreciate helpful comments from T. Maccacaro and R. Della Ceca.
AC acknowledges partial support from the European Community
under the EEC contract No. CHRX-CT92-0033 and Prof. Joachim Tr\"umper
for the hospitality at MPE and for generous support.

\clearpage
\samepage\onecolumn


\newpage\noindent

 {\Large\bf FIGURE CAPTIONS}

\vskip 0.5 true cm
\noindent
{\bf Fig. 1a.} The XRB spectrum compared with our baseline model.
The 0.5-2.0 keV
data are from ROSAT (solid line: Hasinger 1992), while the high energy
data are a
compilation of best results from various experiments (Gruber 1992).
The solid line represents our best fit, the dotted line represents the
contribution of unabsorbed sources, while the absorbed sources are indicated
by dot-dashed lines. The labels are the
logarithms of the corresponding hydrogen column densities.

\noindent
{\bf Fig. 1b.} Percentage deviations from the best fit analytical
approximation (Gruber 1992).
The error bars represent typical one sigma measurement errors
at various energies.

\noindent
{\bf Fig. 2.} The computed predicted counts (solid line) in the soft
(0.5-2.0 keV) band compared with the EMSS AGN counts (dashed lines, Della Ceca
et al. 1992) the ROSAT counts (H93)
and ROSAT fluctuation analysis (dashed area, H93). The predicted contribution
of the unabsorbed sources is shown with a dotted line, while the absorbed
sources are represented by dot-dashed lines
(the labels are the logarithms of the corresponding hydrogen
column densities).

\noindent
{\bf Fig. 3.}
The EMSS AGNs redshift distribution (solid histogram) compared with the
model prediction (dotted histogram).

\noindent
{\bf Fig. 4.}
The ROSAT redshift distribution of the so far identified AGNs in five
deep ROSAT fields with $S(0.5-2.0\ keV)\ \gsimeq\ 10^{-14}$ \ergcms $\ $
(solid histogram) compared with the model predicted distribution
(dotted line).

\noindent
{\bf Fig. 5.} The computed predicted counts in the hard (2-10 keV) energy
range (solid line) compared with the GINGA fluctuation analysis results
(dashed area) and
with the HEA-1 A-2 AGNs surface density at $\sim 3 \cdot 10^{-11}$ ergs
cm$^{-2}$ s$^{-1}$. The predicted contribution of the unabsorbed
sources is shown with a dotted line, while the contribution of absorbed
sources is represented by dot-dashed lines (the labels are the
logarithms of of the corresponding column densities).
The upper dashed line represents the summed contribution
of AGNs and Clusters of Galaxies.

\noindent
{\bf Fig. 6.}
The HEAO-1 A-2 (Piccinotti et al. 1982) AGN redshift distribution (solid
histogram) compared with the model prediction (dotted line).

\noindent
{\bf Fig. 7.}
The model predicted absorption distribution at the HEAO-1 A-2 flux limit
(dotted line) compared with the observed absorption distribution of the
AGNs in the Piccinotti et al. sample (solid line, from Turner \& Pounds 1989).

  \end{document}